# Direct-Coupling Analysis of nucleotide coevolution facilitates RNA secondary and tertiary structure prediction

Eleonora De Leonardis[1,2,3+], Benjamin Lutz[4,5+], Sebastian Ratz[4,5], Simona Cocco[3], Remi Monasson[6], Alexander Schug[4*], Martin Weigt[1,2*]

[1] Computational and Quantitative Biology, Sorbonne Universités, Université Pierre et Marie Curie, UMR 7238, 75006 Paris, France

[2] Computational and Quantitative Biology, CNRS, UMR 7238, 75006 Paris, France

[3] Laboratoire de Physique Statistique de l'Ecole Normale Supérieure, associé au CNRS et à l'Université Pierre et Marie Curie, 75005 Paris, France

[4] Steinbuch Centre for Computing, Karlsruher Institut für Technologie, 76133 Karlsruhe, Germany

[5] Fakultät für Physik, Karlsruher Institut für Technologie, 76133 Karlsruhe, Germany

[6] Laboratoire de Physique Théorique de l'Ecole Normale Supérieure, associé au CNRS et à l'Université Pierre et Marie Curie, 75005 Paris, France

+The authors wish it to be known that, in their opinion, the first two authors should be regarded as joint First Authors.

* To whom correspondence should be addressed. Tel: +49 (0)721 60826303; Fax: +49 (0)721 60824972; Email: schug@kit.edu; Correspondence may also be addressed to Tel. +33 (0)1.44277368; Fax: +33 (0)1 44277336; Email martin.weigt@upmc.fr

## ABSTRACT

**Despite the biological importance of non-coding RNA, their structural characterization remains challenging. Making use of the rapidly growing sequence databases, we analyze nucleotide coevolution across homologous sequences via Direct-Coupling Analysis to detect nucleotide-nucleotide contacts. For a representative set of riboswitches, we show that the results of Direct-Coupling Analysis in combination with a generalized Nussinov algorithm systematically improve the results of RNA secondary structure prediction beyond traditional covariance approaches based on mutual information. Even more importantly, we show that the results of Direct-Coupling Analysis are enriched in tertiary structure contacts. By integrating these predictions into molecular modeling tools, systematically improved tertiary structure predictions can be obtained, as compared to using secondary structure information alone.**

## INTRODUCTION

Experimental work and genomic sequence analysis have revealed that RNAs have a widespread role inside the cell (1-5). In addition to the transmission of genetic information, non-coding RNAs catalyze biochemical reactions and have a crucial role in a multitude of regulatory processes. While some functional RNA act essentially via their single-stranded information or in the context of RNA-protein complexes, in other cases function is directly tied to three-dimensional (3D) RNA structure (6). Gaining such structural knowledge is important for understanding function. Experimental determination of RNA structure, however, remains challenging. Less than 6% of all structures in the RCBS Protein Data Bank (7) contain RNA. Thanks to advances in sequencing technology, many RNAs have been sequenced in different organisms and classified into homologous families in the Rfam database(8). However, out of the 2450 currently listed families, only 59 (2.4%) possess at least one representative PDB structure, even if 954 (39%) contain more than 100 sequences, and 566 (23%) even more than 200 sequences. Computational structure prediction methods are a promising complementary technique to make use of this huge amount of sequences. Unfortunately, even the most advanced computational methods rarely reach RMSD (root mean square deviation) values below 8-12Å in blind RNA structure prediction (9,10), staying well above the resolutions of 2-3Å reachable in X-ray crystal structures. Typically, computational methods (11-13) struggle with various limitations such as alternative solutions, or require expert human intervention. Topologically complex structure cannot be reliably predicted (14,15). Some of the best results in this direction have been obtained by combining computational with experimental approaches: inter-residue contacts inferred from mutational studies can be introduced into computational models as restraints (16,17).

In the related field of protein structure prediction, recent years have seen significant progress in residue contact prediction by exploiting residue coevolution (18). These novel methods are based on two ideas: (i) Tertiary contacts between two residues (even if possibly distant along the primary sequence) lead to correlated patterns in amino-acid occupation, which can be detected by statistical analysis of large multiple sequence alignments (MSA). (ii) Local correlation measures, like Mutual Information (MI), are confound by transitivity effects: Two positions in contact to a common intermediate residue will be correlated even if not directly coupled by adjacency. The advantage of methods like Direct Coupling Analysis (DCA) (19,20) and related methods (21,22) is their capacity to disentangle such direct and indirect effects in order to infer direct couplings from indirect correlations. This leads to a substantial increase in the accuracy of predicted contact maps, with immediate applications to predicting tertiary and quaternary protein structures as diverse as globular proteins (23,24), protein complexes (25-27), active conformations (28) or membrane proteins (29).

Our aim is to propose an efficient pipeline for secondary and tertiary RNA structure prediction based on a coevolutionary analysis of existing homologous sequences. Covariance models for comparative RNA sequence analysis are well known (30,31): MI has been successfully used to infer base-pairs and to predict secondary-structures (30,32,33). It has been argued, however, that non-canonical base pairs involved in tertiary-structure contacts show much less covariation (35), even if some tertiary contacts have been observed to show significant MI (32,33). Here we show, that these weak signals lead to a significantly increased enrichment in 3D contacts when replacing local covariance analysis with DCA. Using a rigorously selected set of riboswitch families with complex structures, we show that DCA can efficiently be integrated into existing tools for RNA secondary- and tertiary-structure prediction. We demonstrate that combined with a standard approach like the Nussinov algorithm (36), DCA leads to a systematic improvement in secondary structure prediction over MI-based methods. Integrating DCA into Rosetta (37), one of the most successful modeling tools for RNA structure prediction (10), we propose a completely automated tertiary-structure prediction scheme. The results systematically improve over those obtained by Rosetta alone, and are competitive to those obtained by other workflows included in the RNA-puzzle experiment, which partially require additional experimental information or expert human manipulations (9,10).

## MATERIAL AND METHODS

### Inference pipeline

When applied to families of homologous RNA, DCA is able to detect secondary and tertiary structure contacts, which in turn can be integrated into existing tools for RNA structure prediction. To do so, we have developed a dedicated pipeline, illustrated in Fig. 1. Starting with a single RNA sequence

- homologs are identified using the Rfam database;
- coevolutionary analysis by DCA is performed;
- secondary structure is predicted by integrating DCA results into a generalized Nussinov algorithm; or
- tertiary structure is predicted by integrating DCA results into molecular modeling by Rosetta.

This pipeline is completely modular, and improvements in each module – better alignments, more precise signal extraction by DCA, integration of alternative (e.g. experimental) knowledge into molecular modeling etc. – can be easily implemented and will lead to an improved overall performance.

### Selection of RNA families

To test the basic pipeline, we have identified a set of test families. Specifically we concentrated on riboswitches, which have non-trivial structures. To generate a representative set of RNA families for both genomic analysis and structure prediction in our study, we systematically selected riboswitch families from the Rfam database by the following criteria: All riboswitches were included, which

- belong to a family with more than 1000 sequences in the Rfam database to provide sufficient statistics for detecting nucleotide coevolution;
- have a maximum of 100 nucleotides in the PDB structure to enable the efficient application of state-of-the-art molecular-modeling tools like Rosetta;
- have a corresponding complete PDB X-ray diffraction structure (each included residue at last represented by a few atoms) with less than 3 Å resolution to be able to evaluate our approach.

In case of multiple PDB structures fulfilling these criteria, the one of highest resolution is chosen. We find that exactly the 6 families described in Table S1 fulfill the selection criteria; no family meeting the criteria was excluded from our analysis. Interestingly, these riboswitches are common targets of many vigorous studies, e.g. (38-41).

For these families, we have extracted multiple-sequence alignments and consensus secondary structures from the Rfam database (42) (version 11.0). To be applicable for structure prediction of specific RNA sequences, the consensus secondary structures were curated by removing contacts, which do not satisfy base pairing in the specific sequence, and possibly by extending helices by adding base pairs adjacent to the consensus secondary structure, cf. Table S2.

These selection criteria are not to be understood as applicability limits of our approach. As is shown in the *Supplementary Information*, Fig. SI1, the contact information extracted by DCA does not change much as soon as subsamples of at least ~250 sequences are used. The entire pipeline is also applicable to sequences of more than 100 nucleotides, at the cost of growing computational costs in particular in the molecular modeling step.

To demonstrate the wide applicability of our pipeline beyond this test set, we have also analyzed 2 RNA families without known 3D structure, namely the glnA riboswitch family RF01739 and the C4 antisense RNA family RF01695. We provide blind structural predictions of the ten highest scoring clusters as supplementary data (pdb-files), which allow for experimental testing.

**Direct-Coupling Analysis**

In the following we briefly recall the main aspects of DCA (for a more detailed description containing technical details please cf. *Supplementary Information*, where the main steps of (19) are summarized and adapted to the specificities of RNA): DCA aims at modeling the sequence variability in the input MSA via a generalized Potts model (or, equivalently, a Markov random field). It assigns a probability to each aligned sequence ($A_1$,..., $A_L$) of L nucleotides or gaps,

EQ. 1

$$P(A_1, \ldots, A_L) = \frac{1}{Z} exp\{\sum_{i<j} e_{ij}(A_i, A_j) + h_i(A_i)\},$$

where $e_{ij}(A_i, A_j)$ denotes the direct coupling between nucleotide $A_i$ in position *i* and nucleotide $A_j$ in position *j*, and $h_i(A_i)$ is a local bias (field) concerning only the nucleotide present in a single position $i$. The partition function *Z* serves to normalize the probability. Parameters have to be adjusted to reproduce the empirical nucleotide statistics extracted from the MSA:

EQ. 2

$$f_i(A_i) = \sum_{\{A_k|k \neq i\}} P(A_1, \ldots, A_L), \qquad f_{ij}(A_i, A_j) = \sum_{\{A_k|k \neq i,j\}} P(A_1, \ldots, A_L),$$

for all positions *i* and *j* and all nucleotides $A_i$ and $A_j$. Here $f_i(A)$ denotes the frequency of occurrence of nucleotide $A$ in position *i* and $f_{ij}(A, B)$ the fraction of sequences having $A$ in position *i* and $B$ in position *j* in the MSA. Parameters are estimated using the mean-field approach of (19). To be able to rank position pairs according to their coupling strengths, the 5x5-dimensional matrices $e_{ij}$ are compressed into a scalar coupling score called Fapc, cf. the *Supplementary information* for technical details. For proteins, it was shown that this procedure strongly outperforms local correlation measures like the mutual information (MI)

EQ. 3

$$M_{ij} = \sum_{A,B} f_{ij}(A,B) log \frac{f_{ij}(A,B)}{f_i(A)\, f_j(B)}$$

with subsequent average product correction (in the following MIapc) (43).

**Secondary structure prediction**

Coevolution of the secondary structure of RNA molecules is well known (44); the complementarity of Watson-Crick base pairs requires coordinated mutations of paired nucleotides in the course of evolution. It is by now part of state-of-the-art methods for predicting RNA secondary structures (45) and aligning multiple RNA sequences (34). Most methods for secondary structure prediction are inspired by Zuker's algorithm for free-energy minimization (46) and may integrate additional information coming from residue co-variation (47); their reliability depends crucially on experimentally determined thermodynamic parameters. Since we are concerned with evaluating the use of statistical sequence information alone, we follow a simpler procedure. It is based on the Nussinov algorithm (36), originally designed to find the secondary structure maximizing the number of base pairs by dynamical programming, and its generalization by Eddy and Durbin (30) to include MI as a residue covariance measure.

We also used a generalized Nussinov algorithm, but with MI replaced by various types of scores, including MI itself, but also MIapc and the DCA score Fapc (19,20). Minimal hairpin loop lengths are set to three nucleotides. Note that these algorithms are not intended as state-of-the-art secondary structure predictors, but their aim is to assess the influence of using DCA scores instead of local MI-based scores in an otherwise identical algorithm. It might be interesting to integrate DCA scores into state-of-the-art methods like RNAalifold (47). This goes, however, far beyond the scope of our current analysis.

The Nussinov algorithm requires a contact-scoring matrix as input, the variant of Durbin et al. uses simply mutual information, which, due to its strictly positive values, tends to overpair. Furthermore, mapped to a specific RNA sequence, it may pair nucleotides showing strong covariation, but not forming a compatible Watson-Crick or wobble pair. To avoid these two effects, we adopt the following strategy for constructing the pair-scoring matrix for a specific target sequence:

1) The matrix is prefilled with the negative value -1 for all incompatible pairs, possible base pairs get zero pair score.
2) The covariance scores (MI, MIapc or Fapc) are sorted, only the $L$ largest values are maintained, with $L$ being the length of the target RNA sequence. For each of these pairs:
    a) We check if both residues can be mapped onto residues on the target sequence (i.e. they do not correspond to gaps in the aligned sequence).

b) If yes, and if the resulting nucleotides are compatible with base pairing, the pair score in the matrix is replaced by the covariance score.

c) Non-mappable or incompatible residue pairs do not lead to changes in the matrix of pair scores.

This construction guarantees that there are at most $L$ positive entries in the matrix, thereby avoiding overpairing. The secondary structure prediction is compared to the secondary structure of the corresponding PDB structure (assessing all residues, which are structurally resolved in the PDB file and aligned in the Rfam alignment, i.e. for which a statistical prediction can be obtained and evaluated).

The dependence of the accuracy of the predicted secondary structure on the number of retained covariance scores is discussed in the *Supplemental Material*: A potential problem is that the number of actual basepairs scales like $L$ with the sequence length, while the number of potential basepairs (and thus coevolutionary scores) scales as $L^2$. For the prediction using DCA scores, the sensitivity and precision of the Rfam consensus structure (when evaluated against the PDB secondary structure) are reached close to a cutoff of $L$ positive entries for our selection of riboswitches, while the predictions using MI and MIapc remain less accurate at comparable sensitivity, cf. Fig. SI2.

**Tertiary contact enrichment**

To characterize tertiary contact enrichment in our predictions, we have adopted the following procedure:

- We concentrate on tertiary contacts, which are not in trivial spatial vicinity based on their proximity to primary or secondary-structure contacts: From the list of all residue pairs we therefore remove (1) all pairs $(i,j)$ with $|i-j|<5$ and (2) for each secondary-structure basepair $(k,l)$, all 25 pairs $(k\pm\{0,1,2\},l\pm\{0,1,2\})$. The resulting list is ordered according to the DCA scores Fapc (equivalently for MI, MIapc) to predict tertiary-structure contacts.

- Minimum heavy-atom distances are determined from a representative PDB structure, pairs with distances up to 8Å (alternatively 4Å, cf. *Results*) are considered as true positive (TP) predictions.

- A sliding window of size $Y$ starting after the $X$ highest-ranking predictions is considered (i.e. the window contains the predictions of ranks $X+1,\ldots,X+Y$). A p-value for contact enrichment is estimated using a binomial null model:

    - For each value of $X$, we calculate the number $T(X,Y)$ of TP predictions within the window $\{X+1,\ldots,X+Y\}$.

- The binomial null model is based on a randomization of the list of all residue pairs of ranks beyond *X* (i.e. excluding the *X* highest-scoring pairs – contacts already predicted cannot be predicted again). Its TP rate $r_0$ is thus given by the fraction of TP contacts in the entire remaining list of residue pairs.
- The p-value is defined as the probability that this null model achieves at least *T*(*X,Y)* TP predictions within a random i.i.d. sample of size *Y*.

This local p-value is a more stringent enrichment criterion than a global p-value calculated on the entire first *X* predictions, since TP rates are empirically observed to decrease almost monotonously (cf. *Results*). In the following, the sliding-window size *Y* is chosen to equal 10% of all elements of the considered list; this value is a compromise between local resolution (small *Y*) and reliability of the p-value (large *Y*).

**Tertiary Structure Prediction via Rosetta**

For tertiary structure prediction, we follow the general procedures from (48) in Rosetta. First, idealized helices are created based on secondary structure information. In a second step RNA junctions and loop motifs are added to the secondary structure. These elements are combined into full models while considering predicted tertiary contacts as energetic constraints. For these constraints, predicted residue-residue contacts are mapped to atom-to-atom distance constraints between two residues based on (49). Based on these constraints and its internal energy model, Rosetta generates thousands of conformations. These structure predictions are ranked by their score and then clustered with a threshold of 4Å to identify representative conformations for the clusters. The detailed procedures are described in the *Supplement*.

## RESULTS

**Direct-Coupling Analysis improves RNA secondary structure prediction**

A first test of the performance of DCA on RNA families is to use direct-coupling scores as input for secondary-structure prediction with a Nussinov-type dynamic programming algorithm, cf. Methods for details. One might argue that Watson-Crick base pairing induces very strong and direct coupling between nucleotides, so DCA might not be needed for secondary-structure prediction. As is shown in Fig. 2, we find, however, a clear increase in sensitivity (i.e. more secondary-structure contacts are predicted) at almost unchanged precision (i.e. the fraction of true-positive predictions in all predictions), as compared to using MI (for the definition of these quantities in terms of true positives (TP), false positives (FP) and false negatives (FN) see the figure caption). The results of MI can be, as found also in proteins, improved by applying the average product correction (43), but this increase is systematically smaller than the one obtained by DCA. In Fig. 3, resulting secondary structures for DCA and MI are shown against a reference structure based on the 3D structural information in the PDB file (see Fig. SI3 for comparison with MIapc). Note that shared errors, both false positives (pairs of sites linked by a green line) and false negatives (blue-filled base-pairs), mostly appear close to non-aligned regions (grey shadowed bases) between the crystal structure sequence and the Rfam alignment. For those regions we cannot properly define the Nussinov scoring matrix, consequently errors are more likely to occur.

Further more, some of the apparently false predictions of DCA with respect to a given experimental structure may actually point to alignment errors of the specific PDB sequence. A particular case is the third stem of 2gdi (base pairs 15-25, 16-24, 17-23 and loop from 18 to 22). First, false positives are located in a highly gapped region: Only the 59% of the sequences in the alignment do not contain any gap in the six positions occupied by these three base pairs. Further analysis shows that, among the ungapped sequences, a majority of 54% is exclusively and fully compatible with the DCA predicted base pairing, while only 2% is exclusively and fully compatible with the selected PDB structure. 31% of the sequences are both compatible with DCA and with the PDB, all other sequences show partial incompatibilities with both secondary structures. In this context, two nucleotides are called compatible with base pairing if they may form a Watson-Crick or wobble pair.

**Direct-Coupling Analysis detects coevolving tertiary structure contacts**

Can we extract contact information going beyond secondary structure when analyzing large sequence alignments of RNA families? It has been argued that coevolutionary signals of tertiary nucleotide-nucleotide contacts are much weaker than secondary-structure signals (35) and may be indistinguishable from noise in many RNA families.

We observe this picture to change once local coevolution measures (MIapc) are replaced by a global approach (DCA). The curves in Fig. 4 display the fraction of TP predictions in dependence of the total number of predictions (TP+FP), averaged over the 6 RNA families, cf. Fig. SI4 for the specific results of each family. The upper two panels use a strict contact cutoff of 4Å, the lower panels the more permissive threshold of 8Å, which is intended to reflect the variability of RNA structures across the members of the RNA family as compared to the chosen representative structure.

At first sight, the results of MI, MIapc and DCA look very similar, cf. the two left panels. This results from the fact that the strongest signal is actually given by secondary-structure contacts, cf. the light lines in the bottom of the two plots, which show only the fraction of tertiary-structure contacts in the highest-scoring predictions, and which remains close to zero at the beginning. To assess more carefully the DCA performance in predicting tertiary-structure contacts, we therefore removed the secondary structure and neighboring pairs (for each secondary structure pair $(i,j)$ we removed also $\{(i\pm0,1,2;j\pm0,1,2)\}$ from the set of all predictions). As is shown in the right panels of Fig. 4, the remaining DCA predictions contain substantially higher contact fractions as compared to MI and MIapc, but they remain still distant from the best possible prediction represented by the black line, where all contacts are listed before the first non-contact is included.

A consistent improvement of DCA over MI and MIapc is also observed when going to a larger set of RNA families. Fig. SI5 shows results for 15 families, which are all Rfam families with more than 1000 sequences, a PDB structure of better than 3Å resolution and with non-trivial tertiary contacts (simple hairpin loops are excluded), but without restrictions on the sequence length or functional class. Large ribosomal RNA are excluded from the statistics since they are completely different in terms of sequence length and number as compared to all other RNA families.

**Direct-Coupling Analysis facilitates RNA tertiary structure prediction using biomolecular modeling tools**

Rosetta builds ideal helices based on secondary structure information (SSI) and adds loop regions to suggest thousands of structural motifs for larger sequence parts based on a fragment library. All the fragments are put together by joining them in a Monte-Carlo procedure to minimize the internal energy. The predicted tertiary residue-residue contacts between nucleic acids from DCA or MI can be added as atom-based constraints (49) to modify the energy term, i.e. they guide the structural prediction by biasing the scoring energies. In the following, we concentrate on six different sets of structural constraints/ information: I) SSI based on the consensus secondary structure in the MSA of RFAM, II / III) SSI plus 25 resp. 100 highest MIapc ranking nucleotide position pairs, IV / V) SSI plus 25 resp. 100 highest DCA ranking position pairs, VI) SSI plus full tertiary inter-residue contact map. Sets I and VI define reference values for the worst vs. best possible situation (i.e. no vs. full tertiary contact information) for RNA structure prediction based on Rosetta energy scores(50). They show the possible improvement, which can be obtained by adding tertiary-structure constraints to SSI. Sets II-V provide, for contacts predicted with MIapc and DCA, two alternative strategies using many but lower quality contact predictions, as compared to less but higher-quality contact predictions. While the first set might guide structure prediction into spurious structures due to false contact predictions, the second set might miss clusters of native tertiary contacts showing lower coevolutionary signal, cf. the contact maps in Figure SI6 of the Supplement.

We add these inter-residue contact predictions as constraints to RNA structure prediction by Rosetta (37), following the protocols of (48).

We run this protocol for the cases of all six riboswitches. Table 1 lists the RMSD for the structural model with the best Rosetta score, and the minimal RMSD for the best 5 resp. 10 Rosetta scores, assuming that additional information (such as known experimental constraints (51)) would allow selecting the most accurate prediction. In all but one case, adding tertiary contacts (set VI) strongly improves RMSD values over the set providing only SSI (set I), illustrating the value of external tertiary-contact information for molecular modeling by Rosetta.

As a general observation DCA guided predictions outperform SSI and MIapc guided ones. In some cases the accuracy of predictions made with full contact information (set VI) is almost reached, cf. also the overlays of the prediction with the native structure in Fig. 5. For sets II and III, based on MIapc as a local covariance measure, we observe variable quality predictions for the different riboswitches (summary of best results over the first 10 predictions of both set II and set III: mean=11.8Å, max=16.3Å, min=7.5Å). In contrast, for sets IV and V, based on DCA, we observe significantly improved prediction quality over set I, and in most cases also over sets II and III (summary of best results over the first 10 predictions of both sets IV and case V: mean=9.6Å, max=12.4Å, min=6.7Å).

Some specific cases are worth to be noticed: Looking at the contact maps for 2gdi in Fig. SI6, the predictions of MI and DCA seem quite similar, but the predicted structures in Table 1 show a much higher accuracy when using DCA. The reason becomes clear when having a closer look at the contact maps. The MI prediction totally misses three of the clusters of native contacts, while the DCA predictions find all clusters. This shows that very few predictions may have a major impact on the final prediction accuracy, when they add correct distance constraints to regions, which are otherwise unconstrained, thereby reducing substantially the version space of feasible 3D structures. This finding is corroborated by the relatively low accuracy in predicting 3owi despite few false positives. The predicted contact maps are dominated by the secondary structure, and relatively limited tertiary structure information is added by our approach.

On the contrary, 2gis shows a relatively large number of false positives, which are distant to any native secondary or tertiary contacts and which may introduce competing structure predictions by Rosetta. In fact, the observed RMSD values are the least robust in our test set.

Another interesting case is 3vrs, which unveils a problem in Rosetta modeling. The results suggest that DCA outperforms the full contact map (All). We find that there are actually better results for the full contact map predictions (~9Å), but they are not detected by the scoring systems, as can be seen in the scatter plots of the clustering for 3vrs in Fig. SI7. Lowest scores are consistently given to clusters of 12-16Å RMSD. A further limitation to the accuracy of Rosetta even in the case of full contact maps results from the fact that we only provide native residue-residue contact lists, but no information about native atom-atom distances.

Further information can be found in the *Supplement*.

## DISCUSSION

These tests demonstrate that, despite the weak coevolutionary signal induced by tertiary-structure contacts, DCA results can be well integrated into RNA 3D structure prediction to systematically and robustly improve prediction accuracy. For a given alignment (Stockholm or Fasta format) obtaining DCA predictions takes a maximum of a few minutes on a standard desktop. The DCA pipeline is based on three steps, cf. the *Supplement*: Re-weighting has $O(M \times L)$ memory and $O(M^2 \times L)$ time complexity, estimating covariance is $O(M \times L^2)$ in memory and in time, and matrix inversion is $O(L^2)$ in memory and $O(L^3)$ in time, with $L$ being the sequence length, and $M$ the sequence number. Mapping the residue-residue contacts to atom-atom contacts only takes a few seconds ($O(L^2)$ time and memory requirements). Rosetta requires the biggest computational effort. The first step is generating the loop region models. For each examined riboswitch, the generation of about 4000 models for each loop region requires a total amount of about 3 CPU days. For the final Rosetta step of assembling the full riboswitch out of the helical and loop region parts, we spent another 36 CPU days per riboswitch to achieve sufficient sampling (2000-5000 models). The entire procedure is fully automated and does not rely on any human curating of the contact predictions or assembly at any step.

This method is directly applicable to other RNAs beyond the ones analyzed here. To show this, we provide two *blind predictions* for two RNA families: The first RNA family is the glnA motif (RF01739), a glutamine-binding riboswitch. With the second example we leave the riboswitches. The C4 antisense RNA family (RF01695) is found in phages infecting bacteria. Both examples fit the selection criteria on sequence number and length we had before, but no experimental PDB structures are known. A specific member sequence of each RNA family has to be chosen for Rosetta modeling: To minimize the risk of alignment errors, we have selected cases where sequence specific DCA predictions and Rfam consensus secondary structures coincide. The resulting computational PDB models (10 best scoring models for each family, built using SSI and 100 DCA predictions) are included in the *Supplement*, 2D and 3D representations are shown in Fig. 6, contact maps in Fig. SI8.

Last but not least, we want to suggest some future directions on how predictions might be improved. While future work will surely lead to improved Rosetta protocols and/or scoring schemes for predicted tertiary structures, our emphasis here is on the possible improvements of the statistical predictions using DCA.

First, the determination of multiple sequence alignment for RNA is a complicated and not fully solved problem, and the procedures used in producing large Rfam MSAs might still lack accuracy. In *Results* we have already shown an example where alignment errors limit the accuracy of secondary structure prediction. To investigate the influence of the alignment quality on the prediction of tertiary contacts, we have done the following numerical experiment: For the Rfam family RF00010 (Bacterial RNase P class A), we have run DCA on three different MSA: *(I)* the full Rfam MSA consisting of 6397 sequences, *(II)* a high-quality structural alignment for 340 sequences (52), and *(III)* a sub-MSA of the Rfam MSA containing the sequences contained in the structural alignment, cf. Fig. 7. Not surprisingly, the small and limited-quality alignment *(III)* performs worst. However, almost comparable advantages over this alignment are achieved by improving the alignment quality at a fixed number of sequences, or by improving the sampling of the RNA family by the full Rfam alignment. It would be interesting to invest some future work in improving the alignment quality *without* resorting to structural information, to get closer to the accuracy of the structural alignment at large sequence number to combine both advantages.

Even for a given MSA, the predictions of coevolutionary analysis might be improved following integrative approaches inspired, e.g., by (53,54), which use the output of DCA (or the very similar PSICOV(22)) together with other features like predicted secondary-structure information and solvent accessibility as an input to machine learning tools. The potential success of such supervised approaches might be limited in the case of RNA due to the small number of Rfam families with known structural representatives. However, there is evidence that a substantial fraction of the contacts in native RNA structures shows some degree of coevolution. To assess this idea quantitatively, we have calculated a p-value for the enrichment of contacts within the coevolutionary predictions. More precisely, we compare the local TP rate in a sliding window with the TP rate of a random prediction (applied to all pairs not included in higher DCA ranks, i.e. to all contacts that remain to be predicted, cf. *Methods*). Fig. 8 reports sensitivity and precision of the predictions of the different covariance methods, at the point where the p-value exceeds, for the first time, a significance threshold of 0.01. While for MI this happens (on average over the 6 test families) when only about 18% of all contacts are included, and at a precision TP/(TP+FP) of only 10%, DCA finds 54% of all contacts at a slightly increased precision of 15%. Again, the application of APC to MI improves the results over simple MI, but results (sensitivity 25%, precision 13%) remain far below those of DCA. A completely random prediction would result in P/(P+N) = 5.7% precision. Consequently, even if DCA reaches a much stronger enrichment in contacts as compared to MIapc, the resulting precision may be too low to be practical for tertiary structure prediction. It seems, however, worth to design filtering methods with the aim of a better discrimination between false and true positive predictions, and to further increase the enrichment of true-positive predictions.

## ACKNOWLEDGEMENT AND FUNDING

The work of SC, RM and MW was partly supported by the Agence Nationale de la Recherche project COEVSTAT [ANR-13-BS04-0012-01]. The work of BL, SR and AS was supported by the Impuls- and Vernetzungsfond of the Helmholtz Association. BL acknowledges hospitality of the UPMC in the initial phase of this work. Funding for open access charge: Karlsruhe Institute of Technology.

## TABLE AND FIGURES LEGENDS

| PDB Rfam | Meff/ L | Secondary only | + 25 MIapc | + 100 MIapc | + 25 DCA | + 100 DCA | + All |
|---|---|---|---|---|---|---|---|
| 1y26 RF00167 | 5.8 | 21.6 \| 19.6 \| 12.6 | 12.3 \| 12.3 \| 12.2 | 12.4 \| 10.2 \| 10.2 | 11.4 \| 8.9 \| 8.9 | 16.2 \| 11.3 \| 11.3 | 7.6 \| 7.6 \| 7.4 |
| 2gdi RF00059 | 32.2 | 16.1 \| 16.1 \| 16.1 | 17.0 \| 16.3 \| 8.3 | 21.5 \| 18.5 \| 17.6 | 7.5 \| 7.5 \| 7.5 | 7.7 \| 6.7 \| 6.7 | 6.3 \| 6.3 \| 6.3 |
| 2gis RF00162 | 10.9 | 20.7 \| 18.4 \| 17.4 | 21.8 \| 21.8 \| 17.5 | 22.2 \| 7.7 \| 7.3 | 22.1 \| 10.5 \| 10.5 | 13.9 \| 13.8 \| 10.6 | 9.9 \| 5.8 \| 5.8 |
| 3irw RF01051 | 11.3 | 17.0 \| 17.0 \| 10.2 | 8.8 \| 7.5 \| 7.5 | 8.5 \| 8.4 \| 7.6 | 9.3 \| 9.2 \| 8.6 | 8.1 \| 7.8 \| 7.8 | 6.5 \| 6.5 \| 6.5 |
| 3owi RF00504 | 20.3 | 13.5 \| 13.5 \| 13.5 | 14.4 \| 14.4 \| 13.3 | 17.1 \| 15.2 \| 14.4 | 14.6 \| 12.4 \| 11.7 | 16.2 \| 12.9 \| 11.6 | 10.0 \| 6.3 \| 6.3 |
| 3vrs RF01734 | 8.3 | 15.8 \| 14.2 \| 12.5 | 15.1 \| 14.6 \| 14.6 | 15.0 \| 14.4 \| 13.4 | 15.6 \| 12.3 \| 12.3 | 11.8 \| 11.8 \| 8.8 | 12.9 \| 12.0 \| 12.0 |

Table 1. Structure predictions based on different residue-residue contact maps for six representative riboswitches. Each entry lists the lowest RMSD (Å) against the experimental structure of the best / best 5 / best 10 scoring clusters of Rosetta predictions. In the second column we show the ratio between the effective number of sequences in the alignment (Meff – defined as the number of sequences with pairwise sequence identities below 90%, cf. the *Supplementary Information* for details) over the length of those sequences including gaps (L). In the third column we use secondary structure information (SSI) only. In the 4th and 6th column we add to SSI atomic level constraints based on the 25 best scoring MIapc / DCA nucleotide pairs, while in the 5th and in the 7th the corresponding 100 best scores are used. In the last column, we base the constraints on the complete native contact map. In every case, atomic constraints are added as soft homogeneous distance constraints, i.e. with unknown native distances. We observe considerable improvement of predictions by adding tertiary constraints based on DCA possessing an edge over MIapc alone.

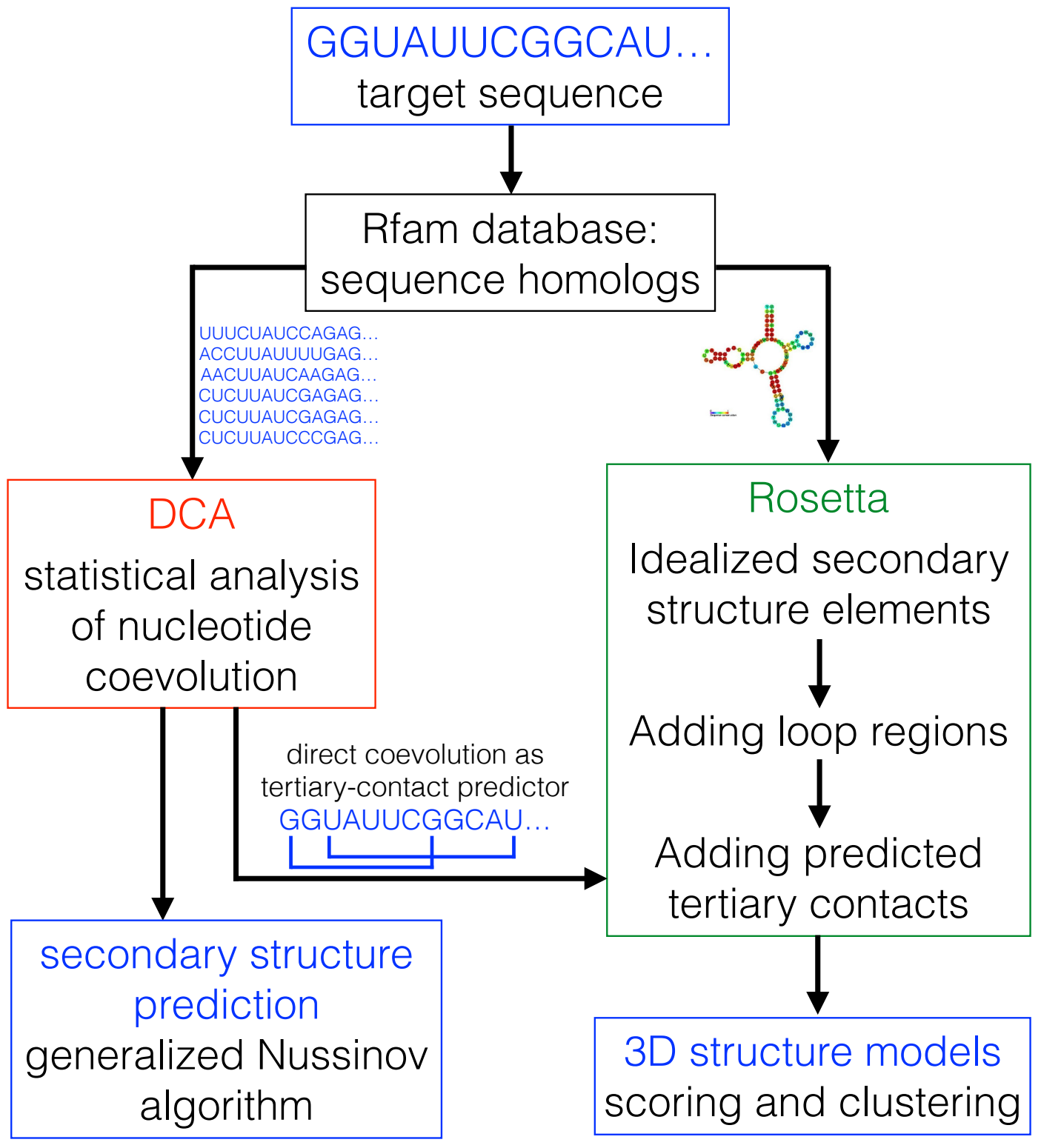

Figure 1. The pipeline for DCA-guided secondary and tertiary structure prediction. Starting from a target sequence, for which structure shall be predicted, we identify sequence homologs using the Rfam database. The corresponding MSA is analyzed by DCA. To predict secondary structure (left side), DCA results are processed by dynamic programming using a generalized Nussinov algorithm. To predict tertiary structure (right side), DCA results are used to predict tertiary structure contacts. The Rfam consensus secondary structure, after curation for the specific target nucleotide sequence, is fed into Rosetta, together with the DCA predictions. Rosetta predictions are clustered according to RMSD, and scored using standard Rosetta energy scores. Representative structures of the best-scoring clusters are returned as the final output of our structure prediction pipeline.

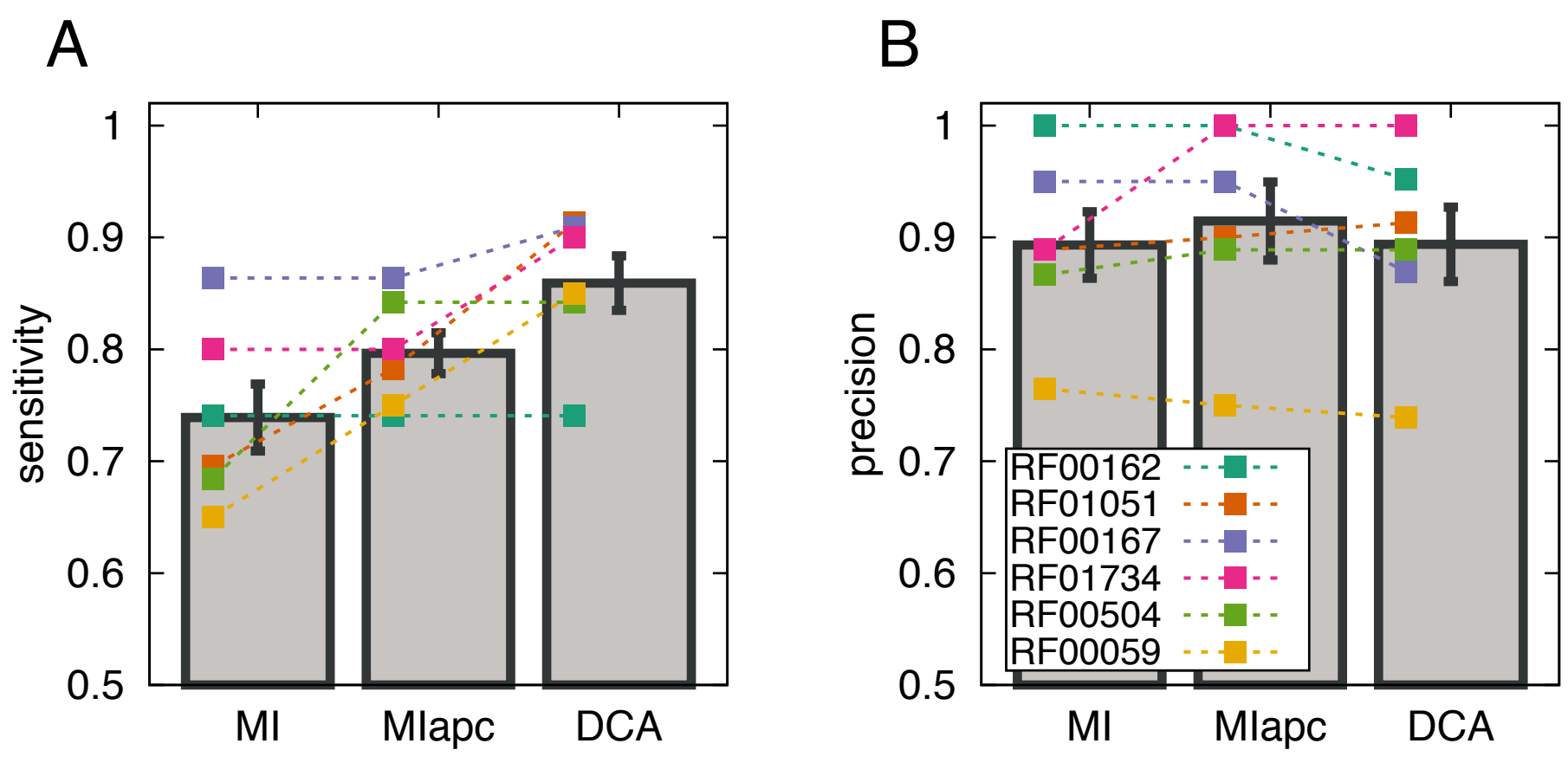

Figure 2. Quality of the secondary structure predicted using the procedure described in Methods, based on mutual information (MI), mutual information with average product correction (MIapc) and the DCA score Fapc: (A) sensitivity TP/(TP+FN), measured as the fraction of predicted contacts (TP) out of all existing contacts (TP+FN); (B) precision TP/(TP+FP), measured by the fraction of predictions, which are true secondary structure contacts. Colored lines show the performance for individual RNA families, grey bars the average performance. DCA is found to consistently increase the sensitivity over both MI scores without losing in precision.

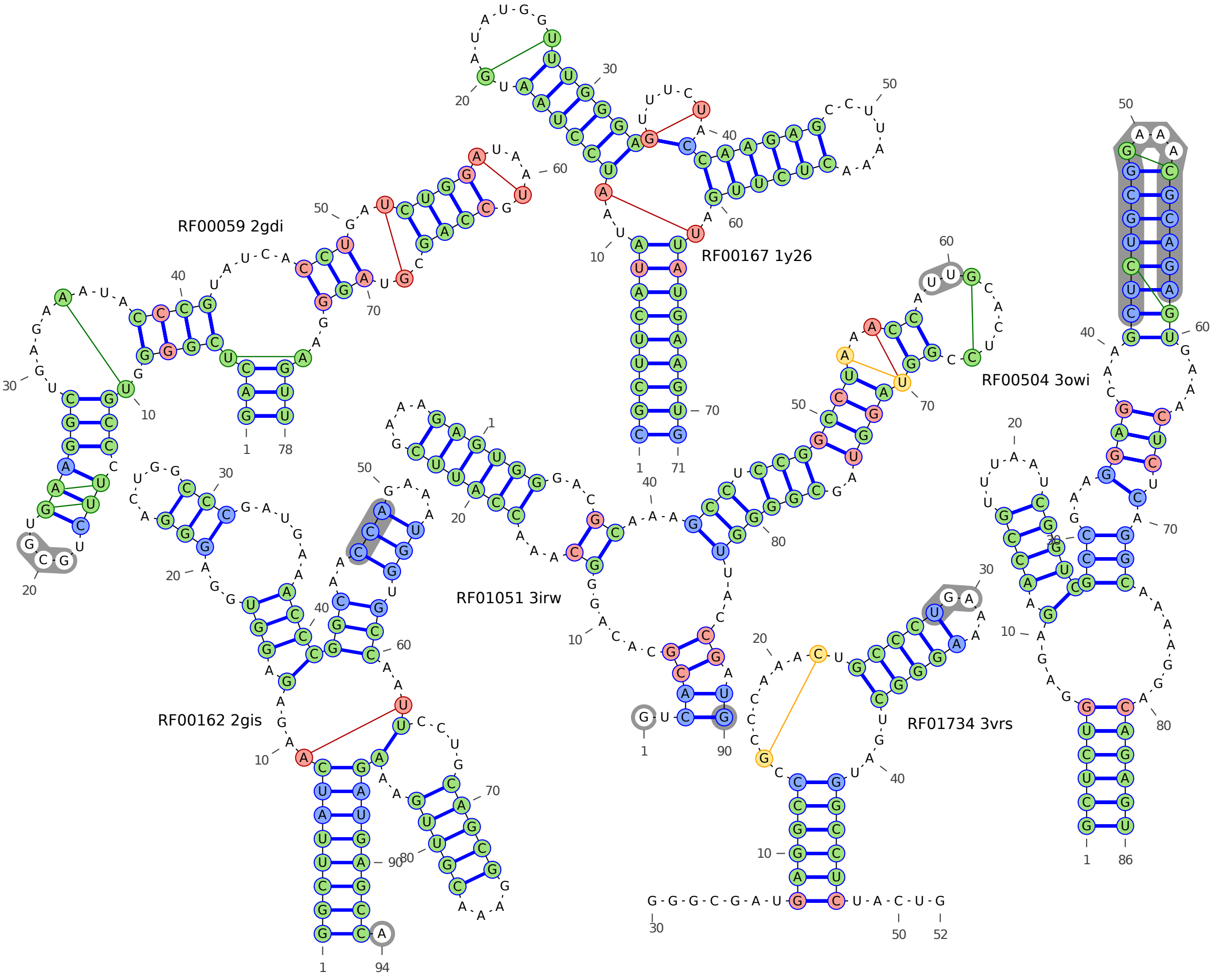

Figure 3. Comparison among PDB secondary structures and predictions using MI and DCA. The underlying secondary structure (blue lines) is derived from base pairs in PDB file. Red-filled base pairs belong exclusively to the DCA predicted structures (DCA TP), yellow-filled exclusively to the MI predicted structures (MI TP), and green-filled ones are found in both the DCA and the MI predictions (both DCA and MI TP). Blue-filled base pairs have not been predicted by coevolutionary analysis (both DCA and MI FN). Lines linking nucleotides outside the secondary structure represent false positives: yellow lines for MI, red lines for DCA and green lines for both. Grey-shaded bases represent non-aligned regions between PDB sequence and Rfam alignment. Basepairs including them cannot be predicted from the Rfam alignment, they have therefore been excluded from the sensitivity and precision values calculated in Fig. 2. The list of WC basepairs in PDB files is extracted with RNAView package.

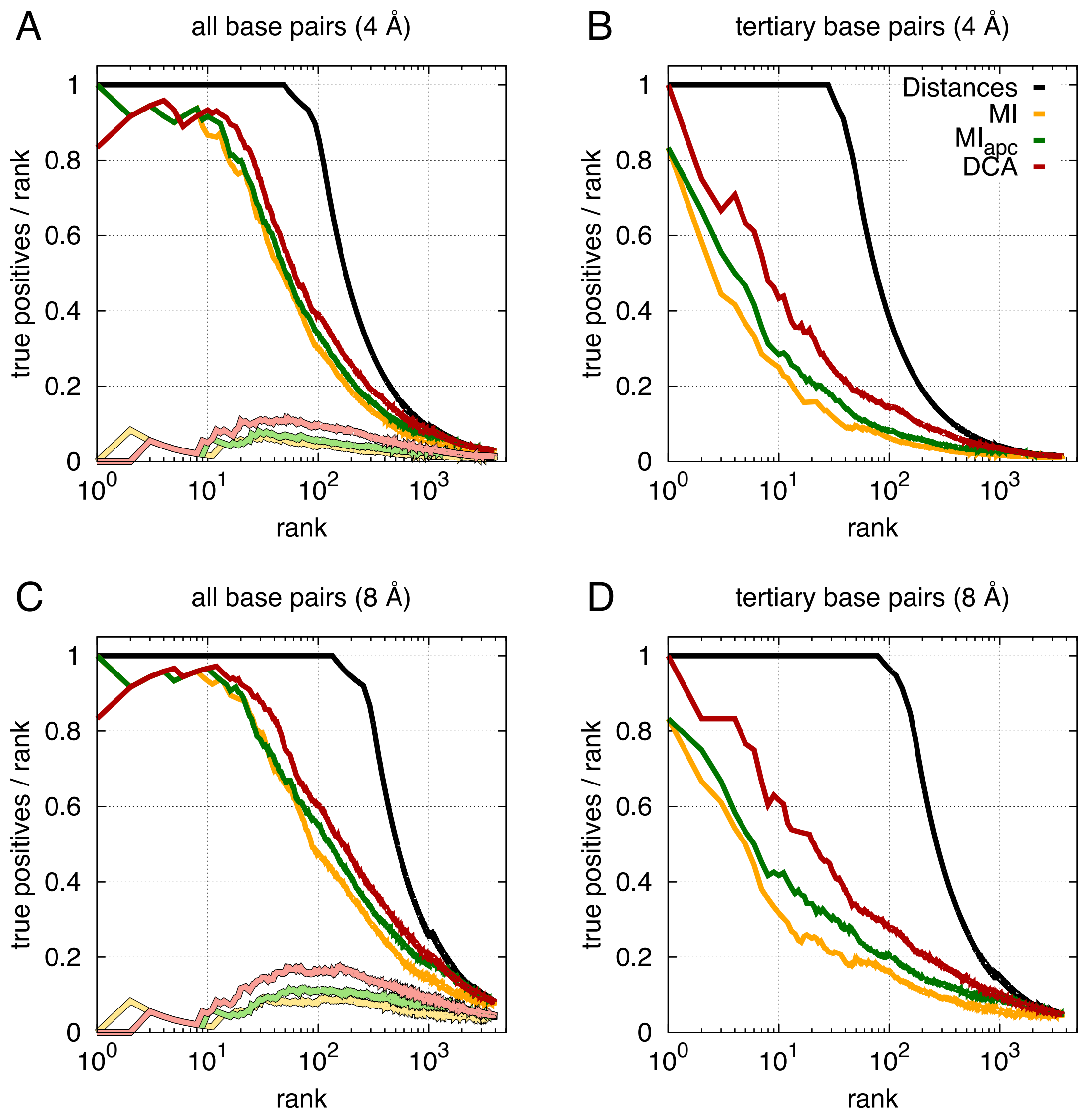

Figure 4. Comparison of the average TP rate (precision TP/(TP+FP)) for contact predictions with DCA (red lines), Mutual Information (MI) (yellow lines) and MIapc (green lines), as a function of the number of predictions (TP+FP). Reference lines (black) are obtained by ranking pairs according to distance, thus they represents the theoretically best possible curves. Two sites are considered to be in contact if the corresponding bases are closer than 4Å (A,B) or 8Å (C,D) in the known crystal structure (distances are measured as minimal distances between heavy atoms). (A,C) All possible pairs (with separation $|i\text{-}j|>4$ along the sequence) are ranked according to their score. Light lines represent the fraction of *tertiary* contacts within these predictions, demonstrating that almost all of the highest DCA scores correspond to secondary-structure base pairs, and tertiary-structure contacts have lower DCA scores. (B,D) Secondary-structure base pairs (and their first and second nearest neighbors) are discarded, only non-trivial 3D contacts are counted. Averages are made over the 6 riboswitches studied, cf. Fig. SI4 for the TP rates of individual families.

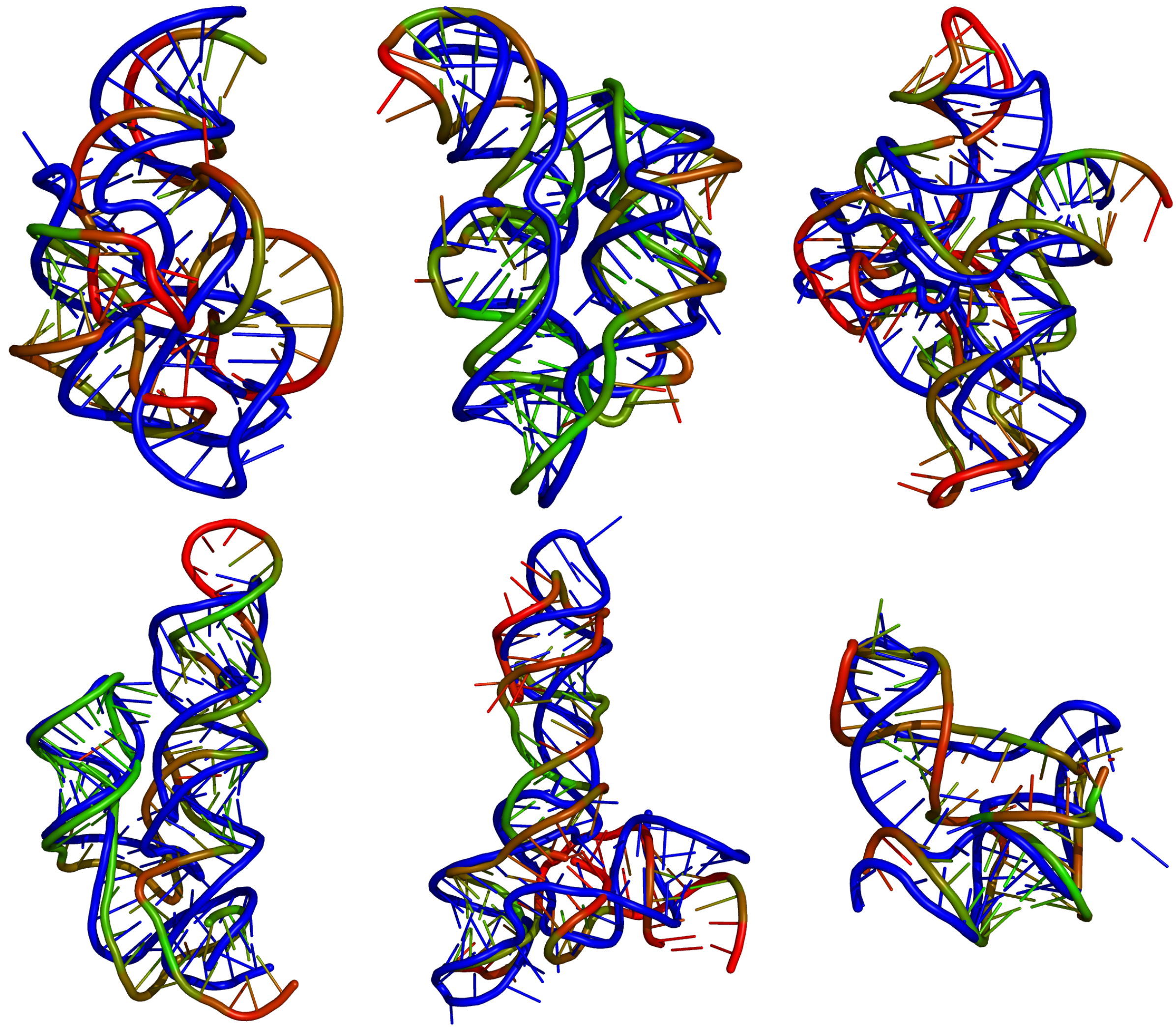

Figure 5. Overlay of the native crystal structures (blue) and structural predictions of the six covered riboswitches. The presented DCA predictions have the lowest native RMSD out of the first 10 clusters (third entry for each column in Table 1). Top row: 1y26 (RF00167) / 2gdi (RF00059) / 2gis (RF00162). Bottom row: 3irw (RF01051) / 3owi (RF00504) / 3vrs (RF01734). The coloring reflects the distance of each residue to the native structure after alignment, ranging from green (0Å, “perfect”) to red (15Å or more, “poor”).

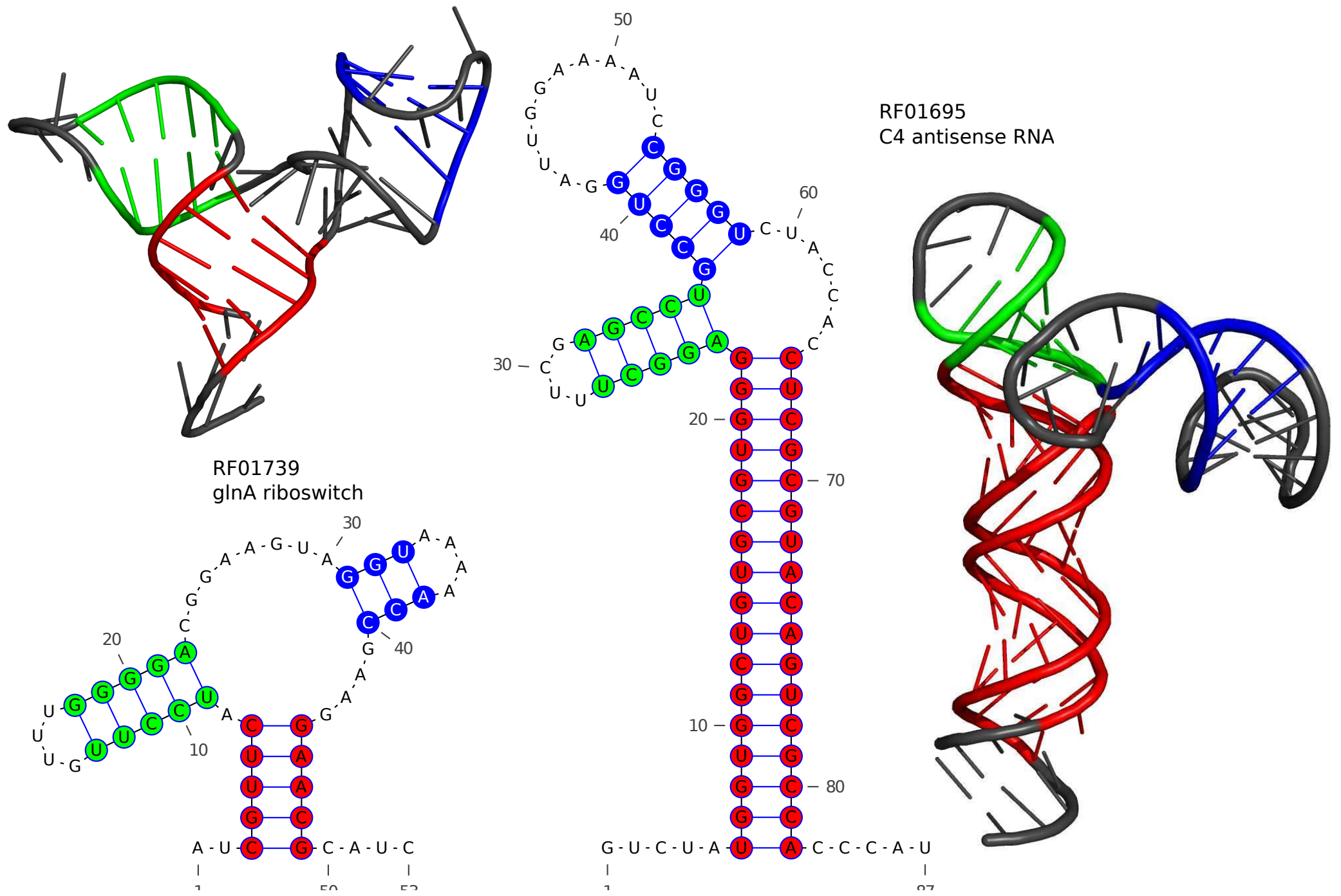

Figure 6. Blind structure predictions for RF01739 (glnA motif) and RF01695 (C4 antisense RNA family), based on secondary-structure information and the 100 highest scoring DCA predictions. The secondary structures have been colored to assure comparability with the tertiary structures, which depict the representative structures of the best-scoring Rosetta clusters. The corresponding supplementary PDB files list the 10 best-scoring clusters for each family.

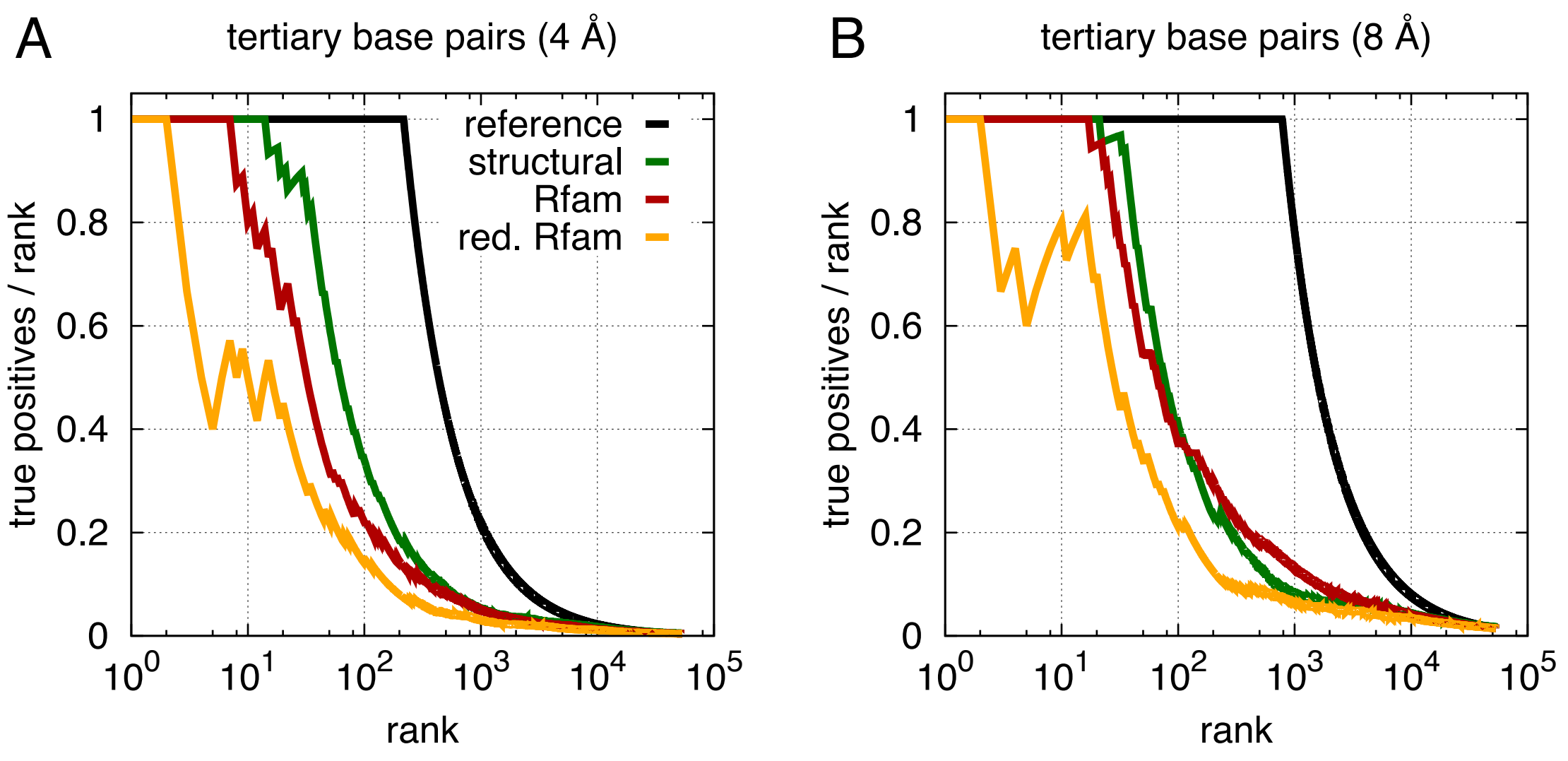

Figure 7. TP rates (precision TP/(TP+FP)) representing DCA performance on three different alignments of sequences of the bacterial Ribonuclease P class A (RF00010), as a function of the number of predictions (TP+FP). (A) refers to a 4Å threshold for contacts, (B) to 8Å. The structural alignment provides the best result in case of a stringent threshold, but it behaves similarly to the complete Rfam MSA if the higher distance threshold is chosen. The reduced Rfam alignment has always the worst performance confirming that both the quality and the size of the alignment are important for DCA.

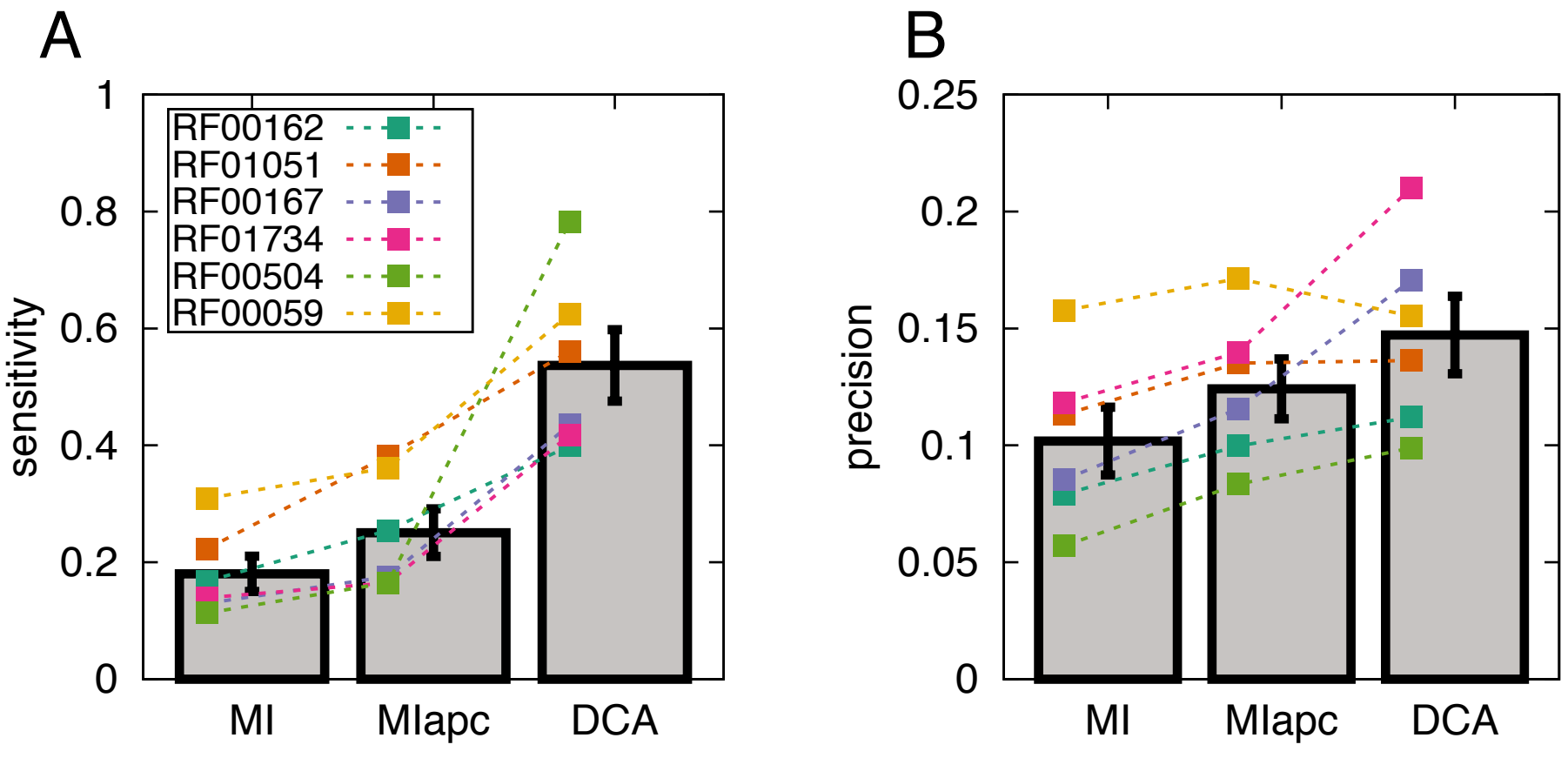

Figure 8. A majority of contacts show coevolution, as detected by statistical enrichment of predicted contacts. Panel (A) shows the sensitivity TP/(TP+FN) of our prediction at the point where the p-value of contact enrichment exceeds 0.01 for the first time. Panel (B) shows the precision TP/(TP+FP) at the same point. Performances for the 6 riboswitch families are shown with colored points, while grey bars display averages. In this picture true contacts are closer than 8Å; only tertiary base pairs are included in the ranking (cf. Figure 4.D, secondary-structure base pairs and their neighbors are excluded). The p-value is computed considering a sliding window containing 10% of all pairs included in the TP rate calculation. This choice was made to take into account the different sequence lengths of different families, as a compromise between resolution and reliability of the p-value.